\documentclass[a4paper]{ESASPCS13Style}
\usepackage{graphicx}
\newcommand{\vtau}[1]{\object{V#1~Tau}}
\newcommand{\ux}{\object{UX~Ari}}
\newcommand{\mm}{\object{MM~Her}}
\newcommand{\rs}{\object{RS~CVn}}
\title{Why do some spotted stars become bluer as they become fainter?}
\author{V. Aarum-Ulv\aa s\inst{1} \and G.\,W. Henry\inst{2}\fnmsep\inst{3}}
\institute{Astrophysikalisches Institut Potsdam -- An der Sternwarte 16,
           D--14482 Potsdam, Germany
           \and
           Center of Excellence in Information Systems, Tennessee State
           University -- 330 10th Avenue North, Nashville, TN~37203, USA
           \and
           Senior Research Associate, Department of Physics and Astronomy,
           Vanderbilt University -- Nashville, TN~37235, USA}

\begin{document}
\maketitle
\begin{abstract}
Chromospherically active, spotted stars generally become
redder as well as fainter when large starspots rotate into
view on the stellar disc.
However, the \rs\ system \ux\
(a triple-lined system), becomes bluer as it gets fainter.
One possible explanation is that hot, bright facular regions
accompany the cool, dark photospheric spots of the active component.
The bluer flux of the hotter, inactive component
does not appear to be sufficient to explain the observed behaviour.
We have begun a search for additional
chromospherically active stars with a similar relation
between colour and brightness, to investigate whether these relations can be
explained in the same way.
Our results for \vtau{711}\ are
presented here, and we conclude that the faculae explanation holds also in
this case.
\end{abstract}

\section{Introduction}
Very active, late-type stars, e.g.\ \ux, \vtau{711}\ and
\mm, show a bluer $B-V$ photometric colour with fainter $V$
photometric magnitude (\cite{crfc96,tei99,auh03});
opposite to what one would expect from spotted stars.
\cite*{amado03} found that active giants (from chromospherically active
single-lined spectroscopic binaries) later than G8 have a bluer $B-V$ than
inactive giants of the same spectral type.
The effect on $B-V$ is smaller than that on $U-B$ reported by \cite*{ab97}.
The most probable explanation, according to both papers, is a facular
component in the photosphere of the active star.
\cite*{aue03} modelled $B-V$ as function of $V$ for \ux\ and
concluded that the relation cannot be explained by the bluer flux of the
hotter secondary component becoming more dominant as the starspots rotate into
view.
It can, however, easily be explained by facular regions surrounding the
starspots.

Our goal is to investigate whether the faculae explanation can apply to other
stars showing the same colour-brightness relation as \ux.
We describe here a similar modelling of the relation between colour
and brightness for \vtau{711}.

\vtau{711} (\object{HR~1099}, \object{HD~22468}) is one of the most active and
well studied \rs\ stars.
The spectroscopic binary star is the primary (A) component of the visual
binary ADS~2644 and
consists of a K1\,IV primary component and a G5\,V secondary component.
The secondary (B) component of the visual binary is situated $6\arcsec$ away
and has spectral classification K3\,V
(\cite{jvdb63,wilson63,wilson64,bf76,fekel83}).

In the following discussion we follow the notation of \cite*{fekel83}:
Components~Aa and Ab refer, respectively, to the more active K1\,IV primary
component and the less active G5\,V secondary component of the spectroscopic
binary.
% Component~Aa refers to the more active K1\,IV primary component of the
% spectroscopic binary.
% Component~Ab refers to the less active G5\,V secondary component of the
% spectroscopic binary.
Component~A refers to the primary component of the visual binary and will also
be used for any combined properties of Aa and Ab.
Finally, component~B refers to the K3\,V secondary component of the visual
binary.

The photospheric and chromospheric activity is mainly associated with
component~Aa, but also Ab has shown some spot activity
(\cite{garcia-alvarezetal03}, and references therein).
This fact makes \vtau{711} particularly interesting, since it results in a
complicated colour-bright\-ness relation as the activity varies on two stars
simultaneously on a timescale of years.

\section{Observations}
Photometric data of \vtau{711} in the Johnson $B$ and $V$ bands were obtained
with the T3 0.4~m automated photo\-electric telescope (APT) at Fairborn
Observatory\footnote{\tt http://www.fairobs.org/}
in southern Arizona.
About half of the data has previously been published by \cite*{hehh95}, where
descriptions of the observing and reduction procedures can be found.

The observations of \vtau{711} were made with a $55\arcsec$ dia\-phragm,
implying that component~B was also included in the measurements.
For this reason we included B also in our modelling.

\begin{figure*}
\centerline{\includegraphics[width=0.8\textwidth]{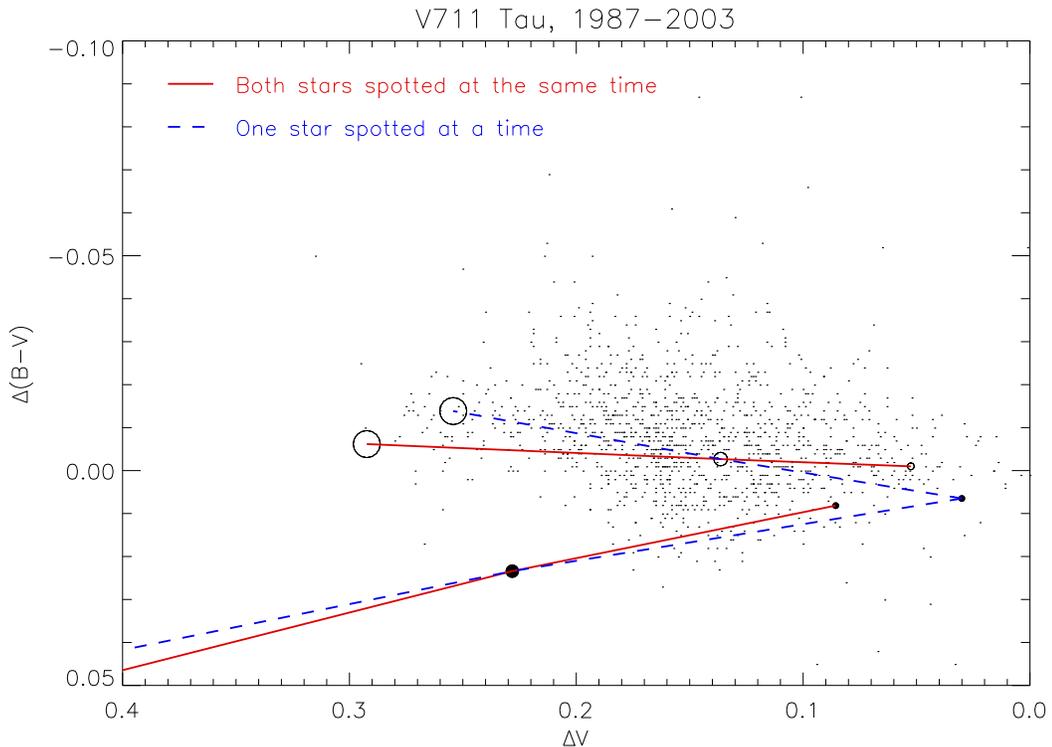}}
\caption{Theoretical calculations of $\Delta V$ and $\Delta(B-V)$ of the
         \vtau{711} system relative to the \vtau{711} system where Aa and Ab
         are both unspotted.
         The observations are shown as dots.
         The filled circles represent models where the active regions
         contain only dark spots.
         The open circles represent models where the Aa active regions
         also contain faculae.}
\label{fig:v711tau}
\end{figure*}

\section{Modelling of spots and faculae}\label{sect:modelling}
We used the method of \cite*{aue03} to calculate the differences in $V$ and
$B-V$ for the whole \vtau{711} system compared to the case of unspotted stars.
The calculations resulted in $\Delta V$ and $\Delta(B-V)$ in the sense spotted
minus unspotted and made use of the stellar and spot parameters as presented
in Table~\ref{tab:v711tau}.
\begin{table}
\caption{Parameters of the \vtau{711}\ system used in our model
         calculations.
         The temperature $T_{\it B}$ and radius $R_{\it B}$ of the visual
         secondary were taken from \protect\cite*{gray92}, given its K3\,V
         spectral classification.
         The facular temperature $T_{\it f}$ was set 250~K higher than the Aa
         effective temperature $T_{\it Aa}$, in accordance with
         \protect\cite*{aue03}.
         Only Aa was assumed to have faculae, in agreement with
         findings by \protect\cite*{ab97} and \protect\cite*{amado03}.}
\label{tab:v711tau}
\begin{center}
\leavevmode
\footnotesize
\begin{tabular}{ll}
\hline\\[-5pt]
Parameter&Reference\\[+5pt]
\hline\\[-5pt]
$T_{\rm Aa}=4800$~K&\cite*{garcia-alvarezetal03}\\
$T_{\rm Ab}=5400$~K&\cite*{garcia-alvarezetal03}\\
$T_{\rm B}=4925$~K&\cite*{gray92}\\
$R_{\rm Aa}=3.3R_\odot$&\cite*{garcia-alvarezetal03}\\
$R_{\rm Ab}=1.1R_\odot$&\cite*{garcia-alvarezetal03}\\
$R_{\rm B}=0.73R_\odot$&\cite*{gray92}\\
$T_{\rm s}=3800$~K&\cite*{garcia-alvarezetal03}\\
$T_{\rm f}=5050$~K&\cite*{aue03}\\
\hline\\
\end{tabular}
\end{center}
\end{table}

Most of the parameter values were taken from \cite*{garcia-alvarezetal03}.
This is the case for the stellar effective temperatures $T_{\rm Aa}$ and
$T_{\rm Ab}$, the
stellar radii $R_{\rm Aa}$ and $R_{\rm Ab}$, and the spot temperature
$T_{\rm s}$.
We used the same $T_{\rm s}$ for both Aa and Ab, in
agreement with \cite*{garcia-alvarezetal03}.
Only the very active Aa was assumed to have faculae, in agreement with
findings by \cite*{ab97} and \cite*{amado03}.
The facular temperature $T_{\rm f}$ was set 250~K higher than the Aa effective
temperature, consistent with the modelling by \cite*{aue03}.

The changes in $V$ and $B-V$ for the whole system were modelled in two cases:
\begin{enumerate}
\item The active regions were visible on both the active stars at the same
      time.
\item The active regions were visible on only one active star at a time.
\end{enumerate}
The observed changes in $V$ and $B-V$ are the results of the rotations of the
Aa and Ab spotted surfaces.
Our model mimics the rotational modulation by varying the relative areas of
the active regions.

\section{Results and discussion}
In order to compare our modelled $\Delta V$ to the observed
values, we derived $\Delta V$ from the observations in the sense observed
value minus the brightest observed value.
Similarly, we derived $\Delta(B-V)$ in the observations in the sense observed
$B-V$ minus the average $B-V$ of those observations having $\Delta V<0.05$.
The brightest $V$ measurement represents the \vtau{711} system as close to
unspotted as we can get from the data.

Our results are presented in Figure~\ref{fig:v711tau}.
The observations are shown as dots.
The model calculation results are represented by open and filled circles with
different sizes.
The filled circles represent cases where the active regions contained only
dark spots, and no faculae.
The open circles represent cases where the active regions contained both dark
spots and faculae (but only for Aa, as explained in
Section~\ref{sect:modelling}).
The varying sizes of the plotting symbols reflect varying relative areas of
the active regions.
Along the lines connecting the plotting symbols, only the relative areas of
the active regions were varied, mimicing the rotation of the system.
The red, solid lines represent case~1, where the active regions on both stars
were visible at the same time.
The blue, dashed lines represent case~2, where the active regions on only one
star were visible at a time.

Figure~\ref{fig:v711tau} shows that the models with spots only (filled
circles) cannot explain the observed relation between $B-V$ and $V$.
This is the case even when Ab is unspotted as the spots of Aa
rotate into view (blue, dashed line, filled circles).
The bluer flux of Ab is insufficient to compensate for the redder
flux of the starspots, given the parameters in Table~\ref{tab:v711tau}.

\cite*{aue03} made a test in the case of \ux\ to find for which stellar and
spot parameters the bluer flux of the hotter secondary component would
compensate for the redder flux of the starspots.
They found that this requires such a low spot temperature and such a small
primary component radius that the active component would be quite different
from its listed spectral type, and this explanation consequently seems
unlikely.

In both cases with faculae (open circles in Figure~\ref{fig:v711tau}) the
modelling results fall within the observed range in $B-V$, thereby providing
an explanation for the observed relation.
The observations do not facilitate a distinction between cases~1 and 2 (red,
solid lines and blue, dashed lines, respectively), however.

It seems likely that the spots of Ab and the active regions of Aa
are visible at the same orbital phase during some observing seasons and
at opposite phases during other seasons.
This effect is likely to cause the large spread in the observed $B-V$
for all observed values of $V$ in Figure~\ref{fig:v711tau} (also suggested by
the two lines connected by open circles).
We also note that the measurements towards redder $B-V$ in
Figure~\ref{fig:v711tau}
can represent cases where the active regions of Aa contain spots
only, and no faculae, as was also found in the case of \ux\ (\cite{aue03}).

We conclude that photospheric facular areas remain the most plausible
explanation for the observed relation between $B-V$ photometric colour and $V$
photometric brightness for \vtau{711}.
The bluer flux of the hotter, less active component only partly compensates
the redder flux of the cool starspots.

\begin{acknowledgements}
This work is supported by the German Research Foundation under project
number TW9249--DFG~STR~645/1--2.
This research has made use of the SIMBAD database,
operated at CDS, Strasbourg, France.
\end{acknowledgements}

\end{document}